\documentclass[epj]{svjour}
\usepackage{graphics}
\usepackage{amsmath}
\usepackage[dvipsnames]{xcolor}
\begin{document}
\title{Structure of the quartetting ground state of $N=Z$ nuclei}
\author{A.G. Serban\inst{1,2,3} \and D. R. Nichita\inst{1,2,3}\and D. Negrea\inst{2}\and V. V. Baran\inst{1,2}\fnmsep% etc
\thanks{\emph{Corresponding author:} vvbaran@fizica.unibuc.ro}%
}                     % Do not remove
\institute{Faculty of Physics, University of Bucharest,
405 Atomi\c stilor, POB MG-11, Bucharest-M\u agurele, RO-077125, Romania \and ``Horia Hulubei" National Institute of Physics and 
Nuclear Engineering, 30 Reactorului, RO-077125, Bucharest-M\u agurele, Romania \and These authors contributed equally.}
\date{Received: date / Revised version: date}
\abstract{The formal equivalence between the quartetting picture and the symmetry restored BCS picture is established for the ground state correlations induced by the general isovector-isoscalar pairing interaction. Multiple ground state structures compatible with the particle number and isospin symmetries are evaluated. The competition of isovector and isoscalar correlations is discussed for the $N=Z$ nuclei above $^{100}$Sn. }

\PACS{
      {21.60.Gx}{Cluster models} %  \and
     % {PACS-key}{discribing text of that key}
     } % end of PACS codes
 %end of abstract
%
\maketitle
\section{Introduction}
\label{intro}

The presence of a collective deuteron-like condensate in nuclei is still actively investigated after more
than six decades since pairing effects were first considered in nuclear physics \cite{Bohr58}.
Theoretical studies on proton-neutron (pn) pairing
have been widely carried out in the framework of
the mean-field  Hartree-Fock-Bogoliubov approximation (see \cite{frauendorf2014} and references therein). One of the salient
characteristics of these approaches is the spontaneous 
breaking of the isospin and particle number symmetries by the solution wavefunctions. While these symmetry violations provide a way to incorporate nontrivial dynamic correlations \cite{sheikh2019} they are also the subject to large 
fluctuations due to the finiteness of nuclear systems. Furthermore, they usually fail to describe properly the coexistence of the $T = 1$ and $T = 0$ pair fields.

Symmetry preserving approaches, on the other hand, show a strong mixing between isovector and isoscalar pairing correlations. Recently, quartetting models have been successfully employed to describe the correlations induced by the proton–neutron pairing interaction \cite{San12,San12a,San14,San15}, but  also by general two-body interactions \cite{Sam151,Sam152,Sam153,Sam154}, in $N=Z$ nuclei. Here, the basic building blocks are correlated four-body structures known as ``quartets'', constructed out of two protons and two neutrons. The general conclusion is that in the quartet approach the isovector and isoscalar pairing correlations mix significantly in $N=Z$ nuclei, contrary to many mean-field studies, and this behaviour remains manifest also for $N>Z$ systems \cite{Neg18}.

The relationship between the mean field models based on pair condensates and the symmetry preserving quartet models has been investigated for particular cases in Refs. \cite{San12,San12a,Dob98,San09,Dob19}, and recently in Ref. \cite{bridging}, in the more general context of a multi-level isovector pairing model. The particle number and isospin projection  of  a proton-neutron BCS pair condensate state naturally generates the collective quartet structure of the ground state ansatz postulated in Ref. \cite{San12} for $N=Z$ nuclei. 

The aim of this work is to further explore the formal correspondence between the quartetting models and the symmetry restored pair condensate approach in the case of combined isovector and isoscalar pairing interactions. In the next {Section} we provide the basic theoretical tools for the study of the isovector-isoscalar pairing Hamiltonian within the above mentioned symmetry conserving approaches, in section \ref{sec:3} we discuss the implications of the Numerical Results, and in the final section we draw Conclusions.

\section{Theoretical background}
\label{sec:1}
\subsection{Isovector-isoscalar pairing Hamiltonian}
%\label{sec:2}
%as required. Don't forget to give each section and subsection a unique label (see Sect.~\ref{sec:1}).

We investigate $N=Z$ nuclei where the nucleons move in a deformed mean field with axial symmetry and  interact via both an isovector and an isoscalar pairing forces. The corresponding Hamiltonian is of the form

\begin{equation}
\label{hamiltonian}
H = \displaystyle \sum_{i} \varepsilon_i N_i + \sum_{i,j,\tau} V_{ij}^{(T=1)} P_{i,\tau}^{\dagger} P_{j,\tau} + \sum_{i,j} V_{ij}^{(T=0)} D_{i,0}^{\dagger} D_{j,0}~,
\end{equation}
where $\varepsilon_i$ represent the single particle energies and $V^{(T)}_{ij}$ are the pairing matrix elements in the isovector $T=1$ and isoscalar $T=0$ channels. The indices $i,j$ denote the single particle doubly degenerate states, ranging from 1 to the number of levels, $N_{\text{lev}}$. The general isovector pairing interaction is expressed in terms of the noncollective pair operators $P_{i, 1}^{\dagger}=\nu_{i}^{\dagger} {\nu_{\bar{i}}^{\dagger}}$,  $P_{i,-1}^{\dagger}=\pi_{i}^{\dagger} {\pi_{\bar{i}}^{\dagger}}$ and $P_{i, 0}^{\dagger}=(\nu_{i}^{\dagger} {\pi_{\bar{i}}^{\dagger}} +\pi_{i}^{\dagger} \nu_{\bar{i}}^{\dagger}) / \sqrt{2}$. The isoscalar proton-neutron pairing interaction is written in terms of the noncollective proton-neutron pair $D_{i, 0}^{\dagger}=(\nu_{i}^{\dagger} \pi_{\bar{i}}^{\dagger}-\pi_{i}^{\dagger} \nu_{\tilde{i}}^{\dagger}) / \sqrt{2}$ . Here, $\bar{i}$ denotes the time conjugate of the state $i$.  The above mentioned pair operators are built using nucleons in time-reversed axially-deformed states, with a well defined projection of the angular momentum $J$ on the $z$-axis, but not a well defined $J$.  Thus, our Hamiltonian (\ref{hamiltonian}) should not be confused with the spherically-symmetric pairing Hamiltonian with interacting $J=0$ isovector pairs and $J=1$ isoscalar proton–neutron pairs, whose quartetting solution is discussed in Ref. \cite{Sam15_740}.

\subsection{Quartetting in $N=Z$ nuclei}
In Ref. \cite{San15}, the pairing correlation energies are described with high precision within the pair-quartet condensation model (PQCM) using the ground state ansatz
\begin{equation}
\label{pqcm}
|{PQCM}\rangle = [ Q_\text{iv}^{\dagger}(x) + Q_\text{is}^{\dagger}(y)]^{n_q} |{0}\rangle
\end{equation}
where $n_q=(N+Z)/4$ is the number of quartets that may be constructed from the valence $N=Z$ protons and neutrons, $Q_{iv}^{\dagger}(x)$  is the collective quartet operator built out of collective isovector pairs $\Gamma_{\tau}^{\dagger} = \sum_i x_i P_{i,\tau}^{\dagger}$ as $Q_{iv}^{\dagger}(x)  = 2 \Gamma_1^{\dagger}(x) \Gamma_{-1}^{\dagger}(x) - [\Gamma_0^{\dagger}(x)]^2$, and $Q_{is}^{\dagger}(y)$ is the squared collective isoscalar pair $\Delta_{0}^{\dagger} = \sum_i  y_i D_{i,0}^{\dagger}$, $Q_\text{is}^{\dagger}(y) = [\Delta_0^{\dagger}(y)]^2$. Both $Q_{iv}^{\dagger}$ and $Q_{is}^{\dagger}$ are by construction isoscalar quartet operators. In the following, we shall use the short definition of ``isovector quartet'' to denote the isoscalar operator $Q_\text{iv}^{\dagger}$ built from isovector pairs. The $x_i$ and $y_i$ mixing amplitudes are two sets of parameters that define the collectivity of the isovector and isoscalar pairs, and they are computed variationally by the minimization of the expectation value of the Hamiltonian (\ref{hamiltonian}) on the normalized state (\ref{pqcm}). Throughout this work, we use ``quartet condensate'' to denote the state constructed by the repeated application of the quartet operator on the vacuum. We note that quartet condensation in the pairing context is fundamentally different than the quartet condensate of {\it in medium bound states} of 
four fermions as, e.g., alpha particles \cite{Sogo09,toh2001} in finite nuclei and infinite nuclear matter, which may undergo Bose-Einstein condensation. 

The structure of the $|PQCM\rangle$  ansatz was inferred in Ref. \cite{San15} from the exact solution of the Hamiltonian (\ref{hamiltonian}) for a set of degenerate states and for pairing forces of equal strength. It is a generalization of the quartet condensation model (QCM) ansatz used in Ref. \cite{San12} to describe the  isovector pairing correlations in the ground state of $N=Z$ nuclei, $|{QCM}\rangle = [ Q_\text{iv}^{\dagger}(x)]^{n_q} |{0}\rangle$. Remarkably, the QCM approach turns out to be perfectly equivalent to the projected BCS ($\mathcal{P}_{NT}$BCS) approach, involving both particle number and isospin restorations, for the isovector pairing case \cite{bridging}. 

In the combined isovector-isoscalar pairing case however, there is a much larger freedom in constructing a collective quartet ansatz for the ground state. Below, we detail the PQCM/$\mathcal{P}_{NT}$BCS correspondence in this case, and also discuss other possible alternatives for the structure of the quartetting ground state.

\subsection{General collective quartet states}
\label{sec:2.3}

We consider the most general symmetry preserving ansatz constructed out of collective quartets,
\begin{equation}
\label{generalc}
|{\bf{c}} \rangle = \sum_{n=0}^{n_q} c_n \, [Q_{iv}^{\dagger}(x)]^{n} \, [Q_{is}^{\dagger}(y)]^{n_q-n} | 0 \rangle= \sum_{n=0}^{n_q} c_n \,|n,n_q-n\rangle,
\end{equation}
whose structure is defined by the expansion coefficients $c_n$ in the space of collective quartet states 
\begin{equation}
    |m,n\rangle = [Q_{iv}^{\dagger}(x)]^{m} \, [Q_{is}^{\dagger}(y)]^{n} | 0 \rangle~.
\end{equation}
For example, the structure of the PQCM ansatz (\ref{pqcm}) involves binomial expansion coefficients
\begin{equation}
\label{cqcm}
    c_n^{(PQCM)} = \frac{n_q!}{n!\, (n_q-n)!}~,
\end{equation} 
while the choice 
\begin{equation}
\label{cpbcs}
    c_n^{(PBCS)} = \frac{1}{(2n+1)!\, (2n_q-2n)!}
\end{equation} 
corresponds to the particle number and isospin projection of the BCS state (see the next subsection and also Ref. \cite{bridging} for more details),
\begin{equation}
    |BCS\rangle=\exp[\Gamma_0^\dagger(x)]\, \exp[i\Delta_0^\dagger(y)] \, |0\rangle~.
\end{equation} 

The analytical structure of the ground state  ansatz may seem rather different in the PQCM and in the $\mathcal{P}_{NT}$BCS cases. Note, however, that by a suitable rescaling of the mixing amplitudes one may fix the coefficients $c_0=c_{n_q}=1$  to allow for a more sensible comparison (excepting, of course, the particular cases $x=0$ or $y=0$). We obtain, e.g. for $n_q=4$,
\begin{equation}
\label{rescalings}
    \begin{aligned}
        c^{(PQCM)}&=(1,4,6,4,1)~,\\
        \tilde{c}^{(PBCS)}&\approx(1,16,42,21,1)~,\\
    \end{aligned}
\end{equation}
where $\tilde{c}$ denote the expansion coefficients obtained after the rescaling. Generally, the structure of the $\mathcal{P}_{NT}$BCS ansatz presents a stronger contribution from the mixed components with similar numbers of isovector and isoscalar quartets than in the PQCM case. At this stage it should be remarked that all terms are highly overlapping, and that the final conclusions regarding the correlations described by the above combinations can only be drawn after the minimization procedure, which yields different mixing amplitudes for each particular case. We discuss this point in more detail in sec. \ref{sec:3} below, where we consider also the comparison with the pure isovector and pure isoscalar quartetting states
\begin{equation}
\label{iv_is}
    |iv\rangle= [Q_{iv}^{\dagger}(x)]^{n_q}|0\rangle ~~~,~~~ |is\rangle= [Q_{is}^{\dagger}(y)]^{n_q}|0\rangle~,
\end{equation}
discussed in Ref. \cite{San15}, as well as the simple superposition of an isovector and an isoscalar quartet condensate
\begin{equation}
\label{ivis}
    |iv\oplus is\rangle= [Q_{iv}^{\dagger}(x)]^{n_q}|0\rangle+ [Q_{is}^{\dagger}(y)]^{n_q}|0\rangle~.
\end{equation}

\subsection{Collective quartets from projected BCS}
\label{sec:2.4}

From a computational perspective, an efficient way to generate an arbitrary collective quartet state $|m,n\rangle = [Q_{iv}^{\dagger}(x)]^{m} \, [Q_{is}^{\dagger}(y)]^{n} | 0 \rangle$ is to project the isospin and the particle number from a pair coherent state. In particular, the isoscalar pair coherent state generates upon particle number projection the collective quartet state
\begin{equation}
\label{expis}
    \hat{\mathcal{P}}_{2n} \exp[\Delta_0^{\dagger}(y)] = \frac{1}{(2n)!} [ \Delta_0^{\dagger}(y) ]^{2n} = \frac{1}{(2n)!} [ Q_{is}^{\dagger} (y)]^{n}~.
\end{equation}
For the isovector part, we consider the rotated pair operators $p_{k,1}^{\dagger} = (P_{k,1}^{\dagger} + P_{k,-1}^{\dagger}) /\sqrt{2}$, $p_{k,-1}^{\dagger} =i( P_{k,1}^{\dagger} - P_{k,-1}^{\dagger})/{\sqrt{2}}$, $p_{k,3}^{\dagger} = i P_{k,0}^{\dagger}$, and the corresponding triplet of collective pairs $\gamma_{a}^{\dagger}(x) =  \sum_{j} x_{j} p_{j,a}^{\dagger}$. The integral over all directions in isospin space of the $\gamma$-coherent state is shown in Ref. \cite{bridging} to generate the collective quartet state,
\begin{equation}
\label{expiv}
\int_{S^2} \text{d}\hat{n} \exp(\hat{n} \cdot \vec{\gamma}^{\dagger}) =  \sum_m \frac{1}{(2m+1)!} [ Q_{iv}^{\dagger}(x)]^m~.
\end{equation}

After combining the isovector and isoscalar coherent states, expanding them as BCS-like products and implementing both the isospin and particle number projections,  we obtain 

\begin{equation}
\begin{aligned}
\label{mn}
&|{m,n}\rangle = [Q_{iv}^{\dagger}(x)]^{m} \, [Q_{is}^{\dagger}(y)]^{n} | 0 \rangle=\frac{(2m+1)!\, (2n)!}{(N_{lev}+1) (2N_{lev}+1)}\\
&\times\sum_{j=1}^{2 N_{lev}+1} \sum_{l=1}^{N_{lev}+1} \exp[-i ( 2n \,\varphi_j + m\, \theta_l)]~ \int_{S^2} d\hat{n}\\
&\times \prod_{k=1}^{N_{lev}}[ 1 + x_{k;l} \, \hat{n} \cdot \vec{p}_k^{\, \dagger} + y_{k;j} D_{k,0}^{\dagger} + ( x_{k;l}^2 + y_{k;j}^2 )\, q_k^{\dagger}/2] |0\rangle~,
\end{aligned}
\end{equation}
where we have defined $\varphi_j = {2 \pi j}/({2N_{lev}+1)}$ and  $\theta_j = {2 \pi j}/(N_{lev}+1)$ and used the shorthand notations $x_{k;l} = x_k e^{i \theta_l/2}$ and $y_{k;j} = y_k e^{i \varphi_j}$. Also, we define  $q_i^{\dagger} =  \nu^{\dagger}_i \nu^{\dagger}_{\bar{i}} \pi^{\dagger}_i \pi^{\dagger}_{\bar{i}} $ to be the quartet operator that fills completely the level $i$. For the particle number projection, we have taken into account that the expansion contains $2N_{lev}+1$ terms for the isoscalar pair coherent state, and $N_{lev}+1$ terms for the isovector pair coherent state of Eq. (\ref{expiv}).

As in Ref. \cite{QBCS}, the computations conveniently reduce to the evaluation of the matrix elements of the various operators $N, P^\dagger P$, $D^\dagger D$, on standard proton-neutron BCS states.  In the present approach however, we obtain a particle-number conserving solution at a lower computational cost. Indeed, only three particle number projection sums are required, whereas in Ref. \cite{QBCS} all four Gaussian integrals need to be performed to ensure the quartet coherent state structure. 

Additionally, the BCS-like structure of the states (\ref{mn}) allows for the angular momentum projection techniques to be smoothly carried over from the single-species pairing case \cite{Fellah73} to the present isovector-isoscalar pairing scenario, as will be explored in future studies.

\section{Numerical results}
\label{sec:3}
In order to compare the various possible choices for the ground state structure discussed in sec. \ref{sec:2.3}, we have performed realistic calculations for $N=Z$ nuclei with  valence nucleons outside the closed cores $^{16}$O, $^{40}$Ca and $^{100}$Sn. Following Ref. \cite{San15}, the single-particle states have been generated using the code $ev8$ \cite{ev8} implementing Skyrme-HF calculations performed for axially deformed mean fields with the force Sly4 \cite{sly4} (we disregard the Coulomb interaction). As the model space for the valence nucleons we consider 10 single-particle levels above the closed nuclear core. The resulting levels are doubly degenerate over the projection of the angular momentum on the $z$-axis and also in isospin. As in Ref. \cite{San15}, we consider a zero range pairing force $V^{T=0,1}\left(\vec{r}_{1}, \vec{r}_{2}\right)=V_{0}^{T=0,1} \delta\left(\vec{r}_{1}-\vec{r}_{2}\right)$ with $V_{0}^{T=1}=465$MeV fm$^3$ and $V_{0}^{T=0}=1.5 V_{0}^{T=1}$.

We present in Table \ref{tab:1} the results for the correlation energy obtained by minimizing the energy function $E(x,y)=\langle \psi(x,y)|H|\psi(x,y)\rangle$/$\langle \psi(x,y)|\psi(x,y)\rangle$ with respect to the mixing amplitudes $x_i$ and $y_i$. The correlation energy is defined here as the difference between the ground state energy in the absence of the pairing interaction and the total energy, i.e. $E_\text{corr}=E(V=0)-E(V)$.

In all cases, the PQCM and the $\mathcal{P}_{NT}$BCS states are numerically extremely close to each other, which is also reflected in the overlaps between the two states, presented in the first column of Table \ref{tab:2}. This confirms the equivalence between the quartet condensation approach and the symmetry restored BCS approach in a realistic case of combined isovector and isoscalar pairing.

\begin{table}[h!]
\caption{Correlation energies calculated with the states $|PQCM\rangle$ of Eq. (\ref{pqcm}), $\mathcal{P}_{NT}|BCS\rangle$ of Eq. (\ref{cpbcs}), $|\text{iv}\oplus \text{is}\rangle$ of Eq. (\ref{ivis}) and with $|\text{iv}\rangle$ and $|\text{is}\rangle$ of Eq. (\ref{iv_is}).}
\label{tab:1}       % Give a unique label
\begin{tabular}{lllllll}
\hline\hline\noalign{\smallskip}
& $|PQCM\rangle$ & $\mathcal{P}_{NT}|BCS\rangle$ & $|iv\oplus is\rangle$ & $|iv\rangle$ & $|is\rangle$  \\
\noalign{\smallskip}\hline\noalign{\smallskip}
$^{20}$Ne & 11.38 & 11.38 & 11.38 & 11.31 & 10.92 \\
$^{24}$Mg & 19.31 & 19.31 & 19.29 & 19.17 & 18.91 \\
$^{28}$Si & 18.74 & 18.74 & 18.74 & 18.72 & 18.54 \\
$^{32}$S & 18.64 & 18.64 & 18.64 & 18.59 & 17.75 \\
\\
$^{44}$Ti & 7.09 & 7.09 & 7.09 & 7.08 & 6.33 \\
$^{48}$Cr & 12.76 & 12.76 & 12.75 & 12.69 & 12.22 \\
$^{52}$Fe & 16.34 & 16.34 & 16.30 & 16.19 & 15.59 \\
$^{56}$Ni & 15.73 & 15.73 & 15.73 & 15.72 & 15.56 \\
\\
$^{104}$Te & 4.53 & 4.53  & 4.53  & 4.49 & 4.02 \\
$^{108}$Xe & 8.03 & 8.03 & 8.02 & 7.96  & 6.73\\
$^{112}$Ba & 9.27 & 9.27 & 9.26 & 9.22 & 7.53\\
$^{116}$Ce & 12.40 & 12.40 & 12.39 & 12.39 & 10.08\\
\noalign{\smallskip}\hline\hline
\end{tabular}
% Or use
%\vspace*{5cm}  % with the correct table height
\end{table}

\begin{table}[h!]
\caption{Overlaps (in percentages) between the $|PQCM\rangle$ state of Eq. (\ref{pqcm}) and the states $\mathcal{P}_{NT}|BCS\rangle$ of Eq. (\ref{cpbcs}), $|\text{iv}\oplus \text{is}\rangle$ of Eq. (\ref{ivis}), $|\text{iv}\rangle$ and $|\text{is}\rangle$ of Eq. (\ref{iv_is}).}
\label{tab:2}       % Give a unique label
\begin{center}
\begin{tabular}{lllll}
\hline\hline\noalign{\smallskip}
& $\mathcal{P}_{NT}|BCS\rangle$ & $|iv\oplus is\rangle$ & $|iv\rangle$ & $|is\rangle$  \\
\noalign{\smallskip}\hline\noalign{\smallskip}
$^{20}$Ne & 100 & 100 & 99.66 & 97.92 \\
$^{24}$Mg & 100 & 99.93 & 99.42 & 98.61 \\
$^{28}$Si & 100 & 100 & 99.94 & 99.28 \\
$^{32}$S & 100 & 99.98 & 99.85 & 97.00 \\
\\
$^{44}$Ti & 100 & 100 & 99.92 & 92.90 \\
$^{48}$Cr & 100 & 99.93 & 99.49 & 96.64 \\
$^{52}$Fe & 99.99 & 99.77 & 98.93 & 95.69 \\
$^{56}$Ni & 100 & 99.98 & 99.95 & 99.17 \\
\\
$^{104}$Te &  100 & 100 & 99.75 & 95.90 \\
$^{108}$Xe & 100 & 99.95 & 99.25 & 83.03\\
$^{112}$Ba & 100 & 99.99 & 99.64 & 79.08\\
$^{116}$Ce & 100 & 99.87 & 99.87 & 67.78\\
\noalign{\smallskip}\hline\hline
\end{tabular}
% Or use
%\vspace*{5cm}  % with the correct table height MERCI!

\end{center}
\end{table}

The excellent agreement of PQCM and $\mathcal{P}_{NT}$BCS is easy to anticipate given their rich and highly entangled structure, containing all possible terms $(Q^\dagger_\text{iv})^m  (Q^\dagger_\text{is})^n$, $0\leq m,n\leq n_q$. It is then interesting to notice from Table \ref{tab:1} that the simpler combination $|\text{iv}\oplus \text{is}\rangle$ of Eq. (\ref{ivis}) is much closer to the full PQCM and $\mathcal{P}_{NT}$BCS than to each of the two isovector and isoscalar condensates taken separately. 

Within the present symmetry conserving approach, the isovector and isoscalar pairing correlations generally coexist and they cannot be easily disentangled, as seen from the large overlaps in Table  \ref{tab:2} (see also Table 1 of Ref. \cite{San15}). One may however assess their relative strength by evaluating separately the energies for the pure isovector and pure isoscalar condensates of Eq. (\ref{iv_is}). This is in contrast with the mean field approach where only one kind of pair condensate, isovector or isoscalar, is found in the ground state of $N=Z$ nuclei \cite{bertsch10,gez11}, despite being allowed to coexist. In particular, Ref. \cite{gez11} predicts a region of dominating isoscalar correlations above $N=Z=60$. 

To explore the possibility of identifying the signatures of strong isoscalar pairing correlations in this region, we have computed the ground state of the $N=Z$ nuclei above $^{100}$Sn up to $n_q=7$, the results being given in Table \ref{tab:3} below.

\begin{table}[h!]
\caption{Correlation energies for the states $|\text{iv}\oplus \text{is}\rangle$ of Eq. (\ref{ivis}), $|\text{iv}\rangle$ and $|\text{is}\rangle$ of Eq. (\ref{iv_is}) for 1 to 7 quartets above $^{100}$Sn, together with the relative error of the $|\text{is}\rangle$ correlation energies with respect to the $|\text{iv}\oplus \text{is}\rangle$ values.}
\label{tab:3}       % Give a unique label
\begin{center}
\begin{tabular}{lllll}
\hline\hline\noalign{\smallskip}
& $|iv\oplus is\rangle$ & $|iv\rangle$ & $|is\rangle$ & $\text{error}_\text{is} (\%)$  \\
\noalign{\smallskip}\hline\noalign{\smallskip}
$^{104}$Te & 4.65 & 4.63 & 3.71 & 20.20 \\
$^{108}$Xe & 8.07 & 8.07 & 6.46 & 27.76\\
$^{112}$Ba & 9.89 & 9.89 & 7.67 & 22.45\\
$^{116}$Ce & 15.20 & 15.20 & 12.18 & 19.87\\
$^{120}$Nd & 17.76 & 17.76 & 13.70 & 22.86\\
$^{124}$Sm & 20.58 & 20.58 & 14.02 & 31.88\\
$^{128}$Pm & 20.93 & 20.93 & 15.74 & 24.80\\
\noalign{\smallskip}\hline\hline
\end{tabular}
% Or use
%\vspace*{5cm}  % with the correct table height MERCI!

\end{center}
\end{table}

For this computation only, we have extended the pairing window to 12 levels, as to avoid filling up the model space in the cases with a large number of quartets. While the pure isovector quartet condensate solution $|\text{iv}\rangle$ is found to agree very well with the fully correlated ansatz for all nuclei, the large errors for the pure isoscalar solution $|\text{is}\rangle$ show no improvement with increasing mass number. Note that, at variance with our deformed computation, Ref. \cite{gez11} uses a spherical approach; the isoscalar solution is favored in this case due to the small spin-orbit splitting given by the accumulation of low-$j$ orbitals near the Fermi surface in the considered nuclear region \cite{frauendorf2014}. More reliable assessments for the competition between the isovector and isoscalar within the present quartetting approach could be given following the restoration of the rotational symmetry, which is computationally feasible in the BCS-like treatment of Sec. \ref{sec:2.4}.

We leave this aspect to be investigated in detail in future works, and return for the remainder of the section to the formal aspects regarding the structure of the quartetting correlations. We note that all  considered ground states are particular instances of the generic ansatz of Eq. (\ref{generalc}). There is no a priori motive for choosing one set of $c_n$ coefficients over another, and in principle they could also be treated as variational parameters in order to  span the whole space of collective quartet states.

Although the already excellent agreement between the PQCM and the exact solutions \cite{San15} does not justify this supplementary computational effort, it is interesting from a purely theoretical standpoint to determine the structure of the maximally correlated collective quartet state. For the simplest nontrivial case, that of two quartets, we are left with only one independent coefficient after rescaling the mixing amplitudes; we write the most general ansatz as
\begin{equation}
\label{2qc}
    |c\rangle=\left([Q^\dagger_\text{iv}(x)]^2+c \, Q^\dagger_\text{iv}(x)\,Q^\dagger_\text{is}(y)+[Q^\dagger_\text{is}(y)]^2\right)|0\rangle~.
\end{equation}

We present in Fig. 1 the correlation energy $E_\text{corr}$ for the nucleus $^{108}$Xe, obtained after the minimization of the energy function $E(x,y)$ for each fixed value of $c$. Strictly speaking, the maximally correlated collective quartet state ($c=2.2$) turns out to be different from both PQCM ($c=2.0$) and $\mathcal{P}_{NT}$BCS $(c=4.47)$. Nevertheless, the correlation energy variation range is extremely small, less than $0.01$MeV, which for all physical purposes makes all collective quartet ansatzes equivalent.

% For one-column wide figures use Hello :-)
\begin{figure}
\resizebox{0.5\textwidth}{!}{%
  \includegraphics{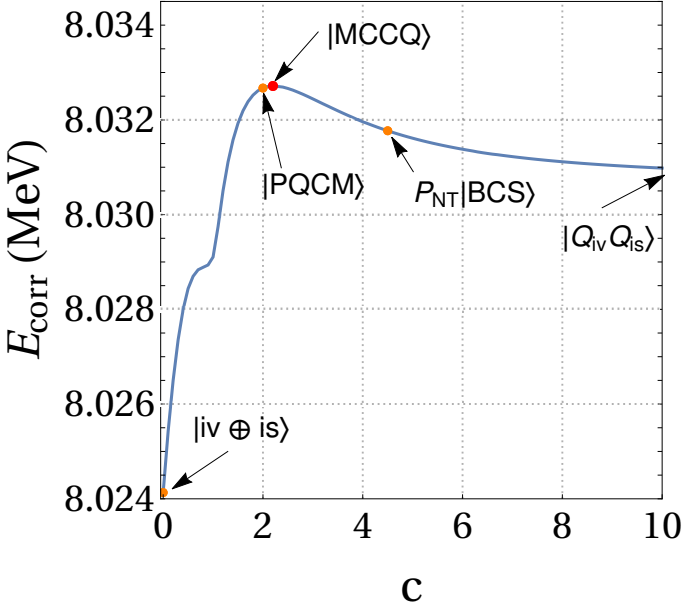}
}
%\vspace{5cm}       % Give the correct figure height in cm
\caption{Correlation energy vs the $c$ coefficient of the $n_q=2$ collective quartet ansatz of Eq. (\ref{2qc}), for the nucleus $^{108}$Xe. The orange dots indicated by arrows refer to the states $|\text{iv}\oplus \text{is}\rangle$ of Eq. (\ref{ivis}) with $c=0$,  $|PQCM\rangle$ of Eq. (\ref{pqcm}) with $c=2$, $\mathcal{P}_{NT}|BCS\rangle$ of Eq. (\ref{cpbcs}) with $c=4.47$. The state $|Q_\text{iv}Q_\text{is}
\rangle=Q_\text{iv}^\dagger(x)Q_\text{is}^\dagger(y)|0\rangle$ corresponds to the limit $c\rightarrow\infty$. The Maximally Correlated Collective Quartet state $|MCCQ\rangle$ with $c=2.2$ (the red dot on the graph) displays the largest correlation energy.}
\label{fig:1}       % Give a unique label
\end{figure}

\section{Summary and Conclusions}

In this work we analyzed the structure of the quartetting ground state of the isovector-isoscalar pairing Hamiltonian for axially-deformed $N = Z$ nuclei. The most general collective quartet state involves all terms built from various numbers of isovector and isoscalar quartet structures. Notable particular examples include the PQCM condensate ansatz and the particle number and isospin projected BCS state. While the functional dependency on the variational amplitudes may differ, the numerical agreement between all collective quartet states is excellent. This is due to the high overlap between the isovector and isoscalar quartet structures, which generally leads to a strong mixing of the two types of correlations in the quartetting approach. This contrasts with the expectation from the mean field picture, where the isovector and isoscalar correlations do not coexist in $N=Z$ nuclei. As opposed to the spherical mean field approach of Ref. \cite{gez11}, the present quartetting approach for deformed nuclei indicates dominating isovector correlations in the region above $N=Z=60$. Full symmetry restoration of the quartetting wavefunction (i.e. including angular momentum projection) is currently under consideration for an improved description of the isovector-isoscalar interplay in realistic scenarios.

\begin{acknowledgement} 
We thank N. Sandulescu, D.  S. Delion and P. Schuck for valuable discussions.
This work was supported by a grant of the Romanian Ministry of Education and Research, CNCS - UEFISCDI,
project number PN-III-P1-1.1-PD-2019-0346, within PNCDI III, and PN-19060101/2019-2022.
\end{acknowledgement}

\bibliographystyle{epj}
\bibliography{epja}
%
% BibTeX users please use
% \bibliographystyle{}
% \bibliography{}
%

\end{document}